\documentclass[sigconf]{acmart}
\def\BibTeX{{\rm B\kern-.05em{\sc i\kern-.025em b}\kern-.08emT\kern-.1667em\lower.7ex\hbox{E}\kern-.125emX}}

\usepackage{booktabs} % For formal tables
\usepackage{float}
\usepackage{url}
\usepackage{natbib}
\usepackage{xspace}
\usepackage{multirow}
\usepackage{amsmath}
\usepackage{todonotes}
\usepackage[utf8]{inputenc}
\usepackage{subcaption}

\newcommand\repname[0]{SPADL\xspace}
\newcommand\repfull[0]{\textbf{S}occer \textbf{P}layer \textbf{A}ction \textbf{D}escription \textbf{L}anguage\xspace}

\newcommand\frameworkname[0]{VAEP\xspace}
\newcommand\frameworkfull[0]{\textbf{V}aluing \textbf{A}ctions by \textbf{E}stimating \textbf{P}robabilities\xspace}

\newtheorem{uaidef}{Definition}

\begin{document}
\fancyhead{}

\title{Actions Speak Louder than Goals:\\ Valuing Player Actions in Soccer}
\renewcommand{\shorttitle}{Actions Speak Louder than Goals: Valuing Player Actions in Soccer}

%\subtitle{Extended Abstract}
%\subtitlenote{The full version of the author's guide is available as
%  \texttt{acmart.pdf} document}

\author{Tom Decroos}
\affiliation{%
	\institution{KU Leuven}
	\streetaddress{Celestijnenlaan 200A, Box 2402}
	\city{Leuven}
	\postcode{3001}
	\country{Belgium}
}
\email{tom.decroos@cs.kuleuven.be}

\author{Lotte Bransen}
\affiliation{%
	\institution{SciSports}
	\streetaddress{Arnhemseweg 2}
	\city{Amersfoort}
	\postcode{3817 CH}
	\country{Netherlands}
}
\email{l.bransen@scisports.com}

\author{Jan Van Haaren}
\affiliation{%
	\institution{SciSports}
	\streetaddress{Arnhemseweg 2}
	\city{Amersfoort}
	\postcode{3817 CH}
	\country{Netherlands}
}
\email{j.vanhaaren@scisports.com}

\author{Jesse Davis}
\affiliation{%
	\institution{KU Leuven}
	\streetaddress{Celestijnenlaan 200A, Box 2402}
	\city{Leuven}
	\postcode{3001}
	\country{Belgium}
}
\email{jesse.davis@cs.kuleuven.be}

%\renewcommand{\shortauthors}{T. Decroos et al.}

%\begin{abstract}
%This paper provides a sample of a \LaTeX\ document which conforms,
%somewhat loosely, to the formatting guidelines for ACM SIG Proceedings.\footnote{This is an abstract footnote}
%\end{abstract}

\begin{abstract}
Assessing the impact of the individual actions performed by soccer players during games is a crucial aspect of the player recruitment process. Unfortunately, most traditional metrics fall short in addressing this task as they either focus on rare actions like shots and goals alone or fail to account for the context in which the actions occurred. This paper introduces (1) a new language for describing individual player actions on the pitch and (2) a framework for valuing any type of player action based on its impact on the game outcome while accounting for the context in which the action happened. By aggregating soccer players' action values, their total offensive and defensive contributions to their team can be quantified. We show how our approach considers relevant contextual information that traditional player evaluation metrics ignore and present a number of use cases related to scouting and playing style characterization in the 2016/2017 and 2017/2018 seasons in Europe's top competitions.
\end{abstract}



%
% Keywords. The author(s) should pick words that accurately describe the work being
% presented. Separate the keywords with commas.

\copyrightyear{2019}
\acmYear{2019}
\setcopyright{acmcopyright}
\acmConference[KDD '19]{The 25th ACM SIGKDD Conference on Knowledge Discovery and Data Mining}{August 4--8, 2019}{Anchorage, AK, USA}
\acmBooktitle{The 25th ACM SIGKDD Conference on Knowledge Discovery and Data Mining (KDD '19), August 4--8, 2019, Anchorage, AK, USA}
\acmPrice{15.00}
\acmDOI{10.1145/3292500.3330758}
\acmISBN{978-1-4503-6201-6/19/08}

\maketitle

\section{Introduction}

How will a soccer player's actions impact his or her team's performances in games? This question is relevant for a variety of tasks within a soccer club such as player acquisition, player evaluation, and scouting. It is also important for the media and building fan engagement, as fans like nothing better than comparing players and arguing why their favorite player is better than the others.

Nevertheless, the task of objectively quantifying the impact of the individual actions performed by soccer players during games remains largely unexplored to date. What complicates the task is the low-scoring and dynamic nature of soccer games. While most actions do not impact the scoreline directly, they often do have important longer-term effects. For example, a long pass from one flank to the other may not immediately lead to a goal but can open up space to set up a goal chance several actions down the line.

Most existing approaches for valuing actions in soccer suffer from three important limitations. First, these approaches largely ignore actions other than goals and shots as most work to date has focused on the concept of the expected value of a goal attempt~\citep{lucey2014quality,caley2015premier,altman2015beyond,mackay2017predicting}. Second, existing approaches tend to assign a fixed value to each action, regardless of the circumstances under which the action was performed. For example, many pass-based metrics treat passes between defenders in the defensive third of the pitch without any pressure whatsoever and passes between attackers in the offensive third under heavy pressure from the opponents similarly. Third, most approaches only consider immediate effects and fail to account for an action's effects a bit further down the line.

To help fill the gap in objectively quantifying player performances, this paper proposes a novel data-driven framework for valuing actions in a soccer game.  Unlike most existing work, it considers all types of actions (e.g., passes, crosses, dribbles, take-ons, and shots) and accounts for the circumstances under which each of these actions happened as well as their possible longer-term effects. Intuitively, an action value reflects the action's expected influence on the scoreline. That is, an action valued at +0.05 is expected to contribute 0.05 goals in favor of the team performing the action, whereas an action valued at -0.05 is expected to yield 0.05 goals for their opponent. Our approach fits within the growing line of data science research for analyzing sports data (e.g.,~\cite{liu:ijcai18,routley:uai15,decroos2018automatic,pappalardo2018playerank}).

In summary, this paper makes the following five contributions:
 \begin{enumerate} 
	\item a language for representing player actions;
	\item a framework for valuing player actions and rating players based on their impact on the game;
	\item a model for predicting short-term scoring and conceding probabilities at any moment in a game;
	\item a number of use cases showcasing our most interesting results and insights;
	\item a Python package\footnote{\url{https://github.com/ML-KULeuven/socceraction}} that (a) converts existing event stream data to our language, (b) implements our framework, and (c) constructs a model that estimates scoring and conceding probabilities.
 \end{enumerate} 

\section{\repname: A language for describing player actions}
\label{sec:representation}

Two primary data sources about soccer games exist that can be used to value actions: (1) event stream data and (2) optical tracking data. Event stream data annotates the times and locations of specific events (e.g., passes, shots, and cards) that occur in a game. Optical tracking data records the locations of the players and the ball at a high frequency using optical tracking systems during games. Multiple different companies (e.g., Opta, Wyscout, STATS, Second Spectrum, SciSports, and StatsBomb) generate one or both of these types of data. Due to the high cost of optical tracking systems, tracking data is only available in wealthy leagues or clubs, while event stream data is more widely and cheaply available. Moreover, optical tracking data is usually not shared across leagues. Hence, this paper focuses exclusively on event stream data. However, the contributions in this paper could also be applied to full tracking data with some minor extensions. A key challenge from a data science perspective is that the nature of the event stream data complicates analysis. We first describe the data science challenges posed by currently available event stream data and then show how our proposed \repname language addresses these challenges.

\subsection{Five data science challenges posed by current event stream data}

The first challenge is that the event stream data serves multiple different objectives (e.g., reporting information to broadcasters, newspapers, or soccer clubs), which means that the data is not necessarily designed to facilitate data analysis. Some important information can be missing (e.g., Wyscout does not record exact end locations for shots). Some of the recorded information might be irrelevant for data analysis and may actually hinder it by increasing the complexity of the preprocessing steps (e.g., Wyscout records duels between two players as two separate events).

The second challenge is that each vendor of event stream data uses their own unique terminology and definitions to describe the events that occur during a game. Hence, software written to analyze data has to be tailored to a specific vendor and cannot be used without modifications to analyze data from another vendor.

The third challenge is that vendors' current event stream formats typically remain backward compatible with their previous formats. Some vendors have been providing data for over a decade, and are unable to alter initial suboptimal design choices. Moreover, what the vendors annotate has evolved and now includes additional events and more detailed information. For example, Opta has four different event types for a shot, depending on its result, which makes it extremely cumbersome to query shot characteristics.

The fourth challenge is that most vendors offer optional information snippets per type of event. For example, for fouls, Opta often specifies more details on the exact type of foul that was committed. While sometimes useful, this dynamic information makes it extremely hard to apply automatic analysis tools.

The final challenge is that most machine learning algorithms require fixed-length feature vectors and cannot handle variable-sized vectors arising from, e.g., the sporadic presense of optional information snippets. Therefore, analysts usually have to write a complicated event preprocessor that extracts the features relevant to their analysis. Developing these preprocessors requires an excessive amount of programming effort and intimate knowledge of the event stream format, but the end result is a one-off script tailored to one specific vendor's current event stream format.

\subsection{Language description}

Based on domain knowledge and feedback from soccer experts, we propose \repname (\repfull) as an attempt to unify the existing event stream formats into a common vocabulary that enables subsequent data analysis. It is designed to be \textit{human-interpretable}, \textit{simple} and \textit{complete} to accurately define and describe actions on the pitch. The human-interpretability allows reasoning about what happens on the pitch and verifying whether the action values correspond to soccer experts' intuitions. The simplicity reduces the chance of making mistakes when automatically processing the language. The completeness enables expressing all the information required to analyze actions in their full context.

To address the challenges posed by the variety of event stream formats and to benefit the data science community, we release a Python package that automatically converts event streams to \repname. Our package currently supports event streams provided by Opta, Wyscout, and StatsBomb. 

\repname is a language for describing player \emph{actions}, as opposed to the formats by commercial vendors that describe \emph{events}. The distinction is that actions are a subset of events that require a player to perform the action. For example, a passing event is an action, whereas an event signifying the end of the game is not an action. We represent a game as a sequence of on-the-ball actions $[a_1,a_2,\ldots,a_m]$, where $m$ is the total number of actions that happened in the game. Each action is a tuple of nine attributes:
 \begin{description} 
	\item[StartTime:] the action's start time,
	\item[EndTime:] the action's end time,
	\item[StartLoc:] the $(x,y)$ location where the action started,
	\item[EndLoc:] the $(x,y)$ location where the action ended,
	\item[Player:] the player who performed the action,
	\item[Team:] the player's team,
	\item[ActionType:] the type of the action (e.g.,\textit{ pass}, \textit{shot}, \textit{dribble}),
	\item[BodyPart:] the player's body part used for the action,
	\item[Result:] the result of the action (e.g., \textit{success} or \textit{fail}).
 \end{description} 

\noindent Note that, unlike all other event stream formats, we always store the same nine attributes for each action. Excluding optional information snippets enables us to more easily apply automatic analysis tools.

We distinguish between 21 possible types of actions including, among others, \textit{passes}, \textit{crossed corners}, \textit{dribbles}, \textit{throw-ins}, \textit{tackles}, \textit{shots}, \textit{penalty shots}, \textit{clearances}, and \textit{keeper saves}. These action types were, in collaboration with domain experts, designed to be interpretable and specific enough to accurately describe what happens on the pitch yet general enough such that similar actions have the same type. The list of all possible action types is in Appendix~\ref{app:action-types}.

We consider up to four different body parts and up to six possible results. The possible body parts are \textit{foot}, \textit{head}, \textit{other}, and \textit{none}. The two most common results are \textit{success} or \textit{fail}, which indicates whether the action had its intended result or not. For example, a pass reaching a teammate or a tackle recovering the ball. The four other possible results are \textit{offside} for passes resulting in an off-side call, \textit{own goal}, \textit{yellow card}, and \textit{red card}.

\section{\frameworkname: A framework for valuing player actions}
\label{sec:framework}
This section introduces the \frameworkname (\frameworkfull) framework for valuing actions performed by soccer players. First, we show how to use scoring and conceding probabilities to compute objective action values. Next, we show how to convert a set of action values to a player rating that represents the player's total offensive and defensive contribution to their team.

\subsection{Converting scoring and conceding probabilities to action values}
Broadly speaking, most actions in a soccer game are performed with the intention of (1) increasing the chance of scoring a goal, or (2) decreasing the chance of conceding a goal. Given that the influence of most actions is temporally limited, one way to assess an action's effect is by calculating how much it alters the chances of both scoring and conceding a goal in the near future. We treat the effect of an action on scoring and conceding separately as these effects may be asymmetric in nature and context dependent.

Suppose that for each game state $S_i=[a_1, \ldots, a_{i}]$, we have access to the probabilities of scoring and conceding in the near future for the home team $h$ and the visiting team $v$. Let $P_{scores}(S_i,h)$ and $P_{concedes}(S_i,h)$ denote the probability of the home team $h$ respectively scoring and conceding in the near future. Similarly, let $P_{scores}(S_i,v)$ and $P_{concedes}(S_i,v)$ denote the probability of the visiting team $v$ respectively scoring and conceding in the near future.

Valuing an action for a team then requires assessing the \emph{change} in probability for both scoring and conceding as a result of action $a_i$ moving the game from state $S_{i-1}$ to state $S_i$. The change in probability for team $x$ scoring, where $x$ can be either the home team $h$ or the visiting team $v$, can be computed as:
\begin{equation}
  \label{eq:hg}
  \Delta P_{scores}(a_i,x)  = P_{scores}(S_i,x) - P_{scores}(S_{i-1},x).
\end{equation}
\noindent This change will be positive if the action increased the probability that team $x$ will score a goal. We call this change $\Delta P_{scores}(a_i,x)$ the \emph{offensive value} of an action $a_i$ for team $x$. Similarly, the change in probability for team $x$ conceding can be computed as:
\begin{equation}
    \label{eq:vg}
	\Delta P_{concedes}(a_i,x)  = P_{concedes}(S_i,x) - P_{concedes}(S_{i-1},x).
\end{equation}
\noindent This change will be positive if the action increased the probability that team $x$ will concede a goal. However, all actions should always aim to decrease the probability of conceding. That is why we call the negation of this change $-\Delta P_{concedes}(a_i,x)$ the \emph{defensive value} of an action $a_i$ for team $x$.

We combine Equations~\ref{eq:hg} and~\ref{eq:vg} to derive an action's total \frameworkname value.
\begin{uaidef}[\frameworkname Value]
\label{def:hattric}
The total \frameworkname value of an action is the sum of that action's offensive value and defensive value.
 \begin{equation}
   \label{eq:hattric}
  V(a_i,x) = \Delta P_{scores}(a_i,x) + (- \Delta P_{concedes}(a_i,x))
 \end{equation}
\end{uaidef}
\noindent Given that we are usually interested in the value of an action for the team of the player performing the action, we use $V(a_i)$ to denote $V(a_i,x_i)$, where $x_i$ is the team of the player performing action $a_i$.

The \frameworkname framework provides a simple approach to valuing actions that is independent of the representation used to describe the actions. The framework's strength is that it transforms the subjective task of valuing an action into the objective task of predicting the likelihood of a future event in a natural way.

\subsection{Converting action values to player ratings}

Our method assigns a value to each individual action. We can aggregate the individual action values into a player rating for multiple time granularities as well as along several different dimensions. A player rating could be derived for any given time frame, where the most natural ones would include a time window within a game, a full game, or a full season. Regardless of the time frame, we compute a player rating in the same manner. Since spending more time on the pitch offers more opportunities to contribute, we compute the player ratings per 90 minutes of game time. Given a time frame $T$ and player $p$, we compute the player's rating as
%\begin{equation}
%	rating(p) = 90 \times \frac{\sum_{a_{i} \in A^{T}_p} V(a_i)}{m},
%\end{equation}
\begin{equation}
rating(p) = \frac{90}{m} \sum_{a_{i} \in A^{T}_p} V(a_i),
\end{equation}
\noindent where $A^{T}_p$ is the set of actions the player $p$ performed during time frame $T$, $V(a_i)$ is computed according to Definition 1, and $m$ is the number of minutes the player played during $T$. This player rating captures the average net goal difference contributed to the player's team per 90 minutes.

%First, players have different positions, and the range of values for the rating may be position dependent. Therefore, comparisons could be done on a per-position basis. Similarly, some players are versatile and what position they play may vary depending on the match. Therefore, it may be interesting to examine a player's rating for each position he or she plays. Second,
Additionally, instead of summing over all actions, a player's rating can be computed per action type. This allows constructing a player profile, which may enable identifying different playing styles. In general, player ratings can be computed along different dimensions, depending on the use case.

\section{Estimating scoring and conceding probabilities}
\label{sec:algorithm}

This section describes our method for estimating the scoring and conceding probabilities required by the \frameworkname framework. Let $goal(h)$ denote a goal scored by  the home team $h$, and $goal(v)$ denote a goal scored by the visiting team $v$. Our task can then be defined as:
 \begin{description} 
	\item[Given:] game state $S_i=[a_1, \ldots, a_{i}]$;
	\item[Estimate:] the probability of scoring and conceding in the near future for the home team $h$ and the visiting team $v$, which we denote by:
	\begin{eqnarray*}
	P_{scores}(S_i,h) &=&  P(goal(h) \in F^k_i | S_i) \\
	P_{concedes}(S_i,h) &=&  P(goal(v) \in F^k_i | S_i)\\
	P_{scores}(S_i,v) &=&  P(goal(v) \in F^k_i | S_i) \\
	P_{concedes}(S_i,v) &=&  P(goal(h) \in F^k_i | S_i)
	\end{eqnarray*}
	where $F^k_i = [a_{i+1}, \ldots, a_{i+k}]$ is the sequence of $k$ actions that follow action $a_i$, and $k$ is a user-defined parameter.
 \end{description} 

\noindent Because $P_{conceding}(S_i,h) = P_{scoring}(S_i,v)$ and $P_{scoring}(S_i,h) = P_{conceding}(S_i,v)$, we only have to estimate the probability of scoring and conceding for one team, and we get the probabilities of the other team for free. We leverage this fact by only estimating the scoring and conceding probabilities for the team that possessed the ball in game state $S_i$. Hence, our task simplifies to two separate binary probabilistic classification problems with identical inputs but different labels.

 \begin{description} 
	\item[Given:] game state $S_i$, where $x_i$ is the team in possession of the ball during $S_i$;
	\item[Estimate:] (1) $P_{scores}(S_i,x_i)$, and (2) $P_{concedes}(S_i,x_i)$.
 \end{description} 

\noindent For both binary classification problems we train a probabilistic classifier to estimate the probabilities. In principle, any machine learning model (e.g., Logistic Regression, Random Forest, or Neural Network) that predicts a probability could be used to address these tasks. However, an important criteria is that the probability estimates should be well-calibrated~\cite{Niculescu-Mizil:icml05}. We use CatBoost~\cite{prokhorenkova2018catboost} and justify this selection empirically in Section~\ref{sec:choice-of-learning-algorithm}.

Applying a standard machine learning algorithm requires converting the sequence of actions $[a_1,a_2,\ldots, a_m]$ describing an entire game into examples in the feature-vector format. Thus, one training example is constructed for each game state $S_i$. We now describe how we compute the labels and features for each game state.

\subsection{Constructing labels}

% For the first classification problem, estimating $P_{scores}(S_i,x_i)$, we assign game state $S_i$ a positive label ($=1$) if the team possessing the ball after action $a_i$ scored a goal in the subsequent $k$ actions. If this did not happen, then the game state is labeled negative ($=0$).

% The labels for the second classification problem, estimating $P_{concedes}(S_i,x_i)$, are constructed in the same way, except that we now check if the team \emph{conceded} a goal in the subsequent $k$ actions after action $a_i$.

For the first classification problem of estimating $P_{scores}(S_i,x_i)$, we assign a game state $S_i$ a positive label ($=1$) if the team possessing the ball after action $a_i$ \emph{scored} a goal in the subsequent $k$ actions, and a negative label ($=0$) in all other cases. Similarly, for the second classification problem of estimating $P_{concedes}(S_i,x_i)$, we assign a game state $S_i$ a positive label ($=1$) if the team possessing the ball after action $a_i$ \emph{conceded} a goal in the subsequent $k$ actions, and a negative label ($=0$) in all other cases.

In both binary classification problems, $k$ is a user-defined parameter that represents how far ahead in the future we look to determine the effect of an action. In this paper, we chose $k=10$ based on domain knowledge and preliminary experiments.

\subsection{Constructing features}

For each example, instead of defining features based on the entire current game state $S_i = [a_1,...,a_i]$, we only consider the previous three actions $[a_{i-2},a_{i-1},a_i]$. Approximating the game state in this manner offers several advantages. First, most machine learning techniques require examples to be described by a fixed number of features. Converting game states with varying numbers of actions, and hence different amounts of information, into this format would necessarily result in a loss of information. Second, considering a small window focuses attention on the most relevant aspects of the current context. The number of actions to consider is a parameter of the approach, and three actions was empirically found to work well. From these three actions, we define features that impact the probability of a goal being scored in the near future.  Based on the \repname representation, we consider three categories of features.

{\bf 1. \repname features.} For each of the three actions, we define a set of categorical and real-valued features based on information explicitly included in the \repname representation. We consider categorical features for an action's type and result, and the body part used by the player performing the action. Similarly, we consider real-valued features for the $(x,y)$-coordinates of the action's start and end locations, and the time elapsed since the start of the game.

{\bf 2. Complex features.} The complex features combine information within an action and across consecutive actions. Within each action, these features include (1) the distance and angle to the goal for both the action's start and end locations, and (2) the distance covered during the action in both the $x$ and $y$ directions. Between two consecutive actions, we compute the distance and elapsed time between them and whether the ball changed possession. These features provide some intuition about the current speed of play.

{\bf 3. Game context features.} The game context features are (1) the number of goals scored in the game by the team possessing the ball after action $a_i$, (2) the number of goals scored in the game by the defending team after action $a_i$, and (3) the goal difference after action $a_i$. We include these features because teams often adapt their playing style to the current scoreline (e.g., a team that is 1-0 ahead will play more defensively than a team that is 0-1 behind).

\section{Experiments}
\label{sec:results}

Evaluating our framework is challenging as no objective ground truth action values or player ratings exist. Therefore, our experiments address three main questions: (1) providing intuitions into how our framework behaves and compares to other metrics, (2) presenting use cases revolving around player acquisition and characterization, and (3) evaluating several of our design decisions.

We focus our analysis on Wyscout data for the English, Spanish, German, Italian, French, Dutch, and Belgian top divisions. We apply the \frameworkname framework to 11565 games played in the 2012/2013 through 2017/2018 seasons. We only consider league games and thus ignore all friendly, cup, and European games.

We train two classification models using the CatBoost algorithm and the feature set detailed in Section~\ref{sec:algorithm} to produce scoring and conceding probabilities, action values, and player ratings. We train the first model on the 2012/2013 through 2015/2016 seasons to produce the outcomes for the 2016/2017 season. Similarly, we train the second model on the 2012/2013 through 2016/2017 seasons to produce the outcomes for the 2017/2018 season.

\subsection{Intuition behind the action values}

\begin{figure}
	%\centering
	%\includegraphics[trim=0.6cm 0 0 0,clip,width=.48\textwidth]{wyscout/messi-phase.pdf}
	\includegraphics[trim=0.55cm 0 0 0,clip,width=.48\textwidth]{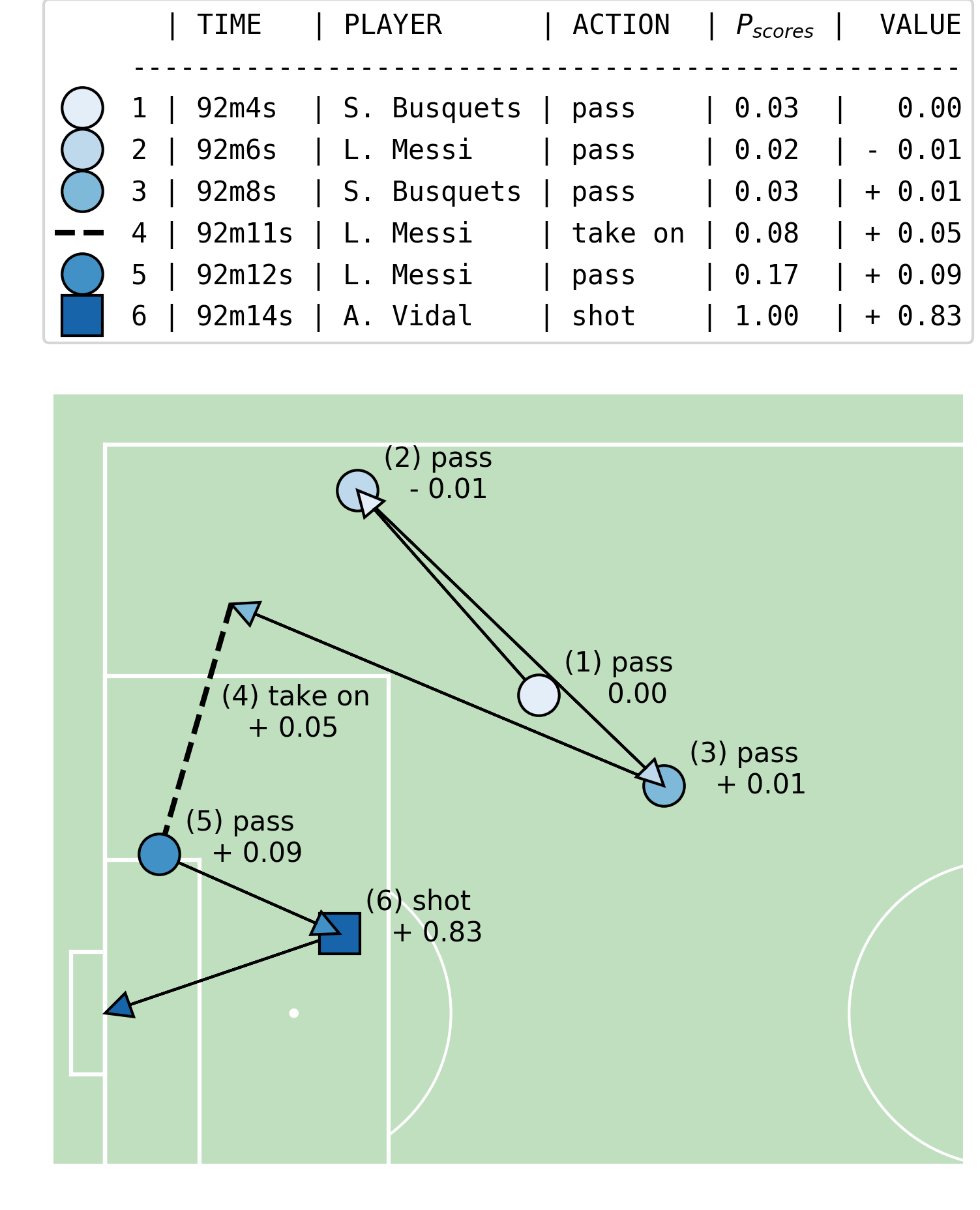}
	\caption{The attack leading up to Barcelona's final goal in their 3-0 win against Real Madrid on December 23, 2017.}
	\label{fig:messi-phase}
\end{figure}
% To illustrate how our framework works, consider the visualization of a failed header from Romelu Lukaku in the 84th minute during Manchester United vs Manchester City on Sunday December 10th 2017 in Figure \ref{fig:lukaku}.

Figure~\ref{fig:messi-phase} illustrates how our framework works by visualizing the actions and their corresponding values that led to Barcelona's goal in the 93rd minute of their game away against Real Madrid on December 23, 2017.

The attack consists of six actions and starts with Sergio Busquets passing the ball towards the right flank (1), which receives a neutral action value of 0.00 since it neither improves nor worsens the situation. The subsequent pass from Lionel Messi back to Busquets (2) is penalized with an action value of -0.01 since it moves the ball backwards to a less favorable position than before. Busquet's excellent through ball to Messi (3), which finally moves the ball closer to the goal, receives an action value of +0.01. Messi receives the ball and dribbles past a Real Madrid defender into the box (4), which receives an action value of +0.05 for significantly raising the scoring odds from 0.03 to 0.08.

Messi's next action showcases his genius, passing the ball backwards and away from the crowded six yard box (5). Our framework awards this pass a value of +0.09 for raising the scoring odds from 0.08 to 0.17. This action shows the power of our framework, which rewards Messi for moving the ball away from the opponent's goal. In a purely data-driven way, our framework identifies this action to be a good choice given the circumstances. To our knowledge, no other method for valuing actions using event stream data would reward Messi for this action. Finally, Aleix Vidal shoots the ball in (6). For converting a 0.17 scoring chance to a goal, our framework rewards Vidal with an action value of +0.83. If Vidal had missed his shot, he would have been penalized with an action value of -0.17.

\subsection{Comparing our \frameworkname player ratings to traditional player performance metrics}

Currently, players' offensive contributions are usually quantified by counting goals and assists, as those events directly influence the score line.\footnote{https://www.squawka.com/en/news/every-player-with-10-goals-and-10-assists-in-europes-top-five-leagues-this-season-ranked-by-contribution-per-90/1031863} Therefore, we compare our \frameworkname player ratings against the following three baseline metrics: goals per 90 minutes, assists per 90 minutes, and goals + assists per 90 minutes. We investigate these metrics' capabilities to identify top players by producing each metric's top-10 list for the 2017/2018 English Premier League season, which are shown in Table~\ref{tbl:player-ratings}. The top 10 in terms of goals per 90 minutes consists of strikers who focus on finishing rather than creating scoring chances. Similarly, the top 10 in terms of assists per 90 minutes mostly consists of midfielders who primarily specialize in setting up chances for their teammates. Furthermore, the ranking in terms of goals + assists per 90 minutes aims to strike a balance between both archetypes.

However, our framework also identifies impactful players who do not rate high on these traditional metrics. First, the top-10 list produced by our \frameworkname framework features Kevin De Bruyne (Manchester City), Eden Hazard (Chelsea), and Riyad Mahrez (Leicester City). Although considered Premier League stars, they do not appear in any of the traditional top-10s. Second, the combined market value for the players in our top-10 list (1,110 million euro) is considerably higher than that for goals (862 million euro), assists (760 million euro), and goals + assists (947 million euro).

These observations suggest that our \frameworkname framework captures players' contributions to their teams' performances better than the traditional player performance metrics.

\begin{table}
	\small
	%\tabcolsep=0.12cm
	%\captionsetup[subtable]{justification=centering}
	\caption{The top-10 players who played at least 900 minutes in the 2017/2018 English Premier League season in terms of (g) goals, (a) assists, (g+a) goals + assists, and (vaep) our \frameworkname player ratings. $R_m$ denotes the rank of the player out of 305 players for metric $m$. The market value denotes the player's market value on February 1, 2019 according to \mbox{Transfermarkt.de}.}
	\label{tbl:player-ratings}
	%\vspace{.3cm}
\begin{subtable}{.48\textwidth}
	\centering
\caption{Top-10 players in terms of goals per 90 minutes (g/90) }
\label{tbl:pr-a}
\begin{tabular}{|clrrr|}
	\toprule
	$R_{g}$ &         Player &   g/90 &  $R_{vaep}$ & Market Value \\
	\midrule
	1 &       M. Salah &  0.986 &           2 &       € 150m \\
	2 &      S. Agüero &  0.960 &          14 &        € 75m \\
	3 &  P. Aubameyang &  0.851 &          42 &        € 75m \\
	4 &        H. Kane &  0.847 &           9 &       € 150m \\
	5 &       G. Jesus &  0.700 &         204 &        € 70m \\
	6 &      O. Niasse &  0.666 &          17 &         € 7m \\
	7 &    R. Sterling &  0.625 &           7 &       € 120m \\
	8 &      C. Austin &  0.612 &         117 &        € 10m \\
	9 &   A. Lacazette &  0.570 &          49 &        € 65m \\
	10 &    P. Coutinho &  0.565 &           1 &       € 140m \\
	\bottomrule
\end{tabular}
\end{subtable}

\begin{subtable}{.48\textwidth}
	\centering
	\vspace{.2cm}
	\caption{Top-10 players in terms of assists per 90 minutes (a/90)}
	\label{tbl:pr-b}
\begin{tabular}{|clrrr|}
	\toprule
$R_{a}$ &         Player &   a/90 &  $R_{vaep}$ & Market Value \\
\midrule
1 &  H. Mkhitaryan &  0.484 &         114 &        € 30m \\
2 &    P. Coutinho &  0.484 &           1 &       € 140m \\
3 &        L. Sané &  0.482 &          47 &       € 100m \\
4 &   K. De Bruyne &  0.467 &           3 &       € 150m \\
5 &       D. Silva &  0.369 &          13 &        € 25m \\
6 &    R. Sterling &  0.347 &           7 &       € 120m \\
7 &  P. Aubameyang &  0.340 &          42 &        € 75m \\
8 &        M. Özil &  0.333 &          15 &        € 35m \\
9 &       P. Pogba &  0.332 &           8 &        € 80m \\
10 &       C. Brunt &  0.327 &          73 &         € 2m \\
\bottomrule
\end{tabular}
\end{subtable}

\begin{subtable}{.48\textwidth}
	\centering
	\vspace{.2cm}
	\caption{Top-10 players in terms of goals + assists per 90 minutes (g+a/90)}
	\label{tbl:pr-c}
\begin{tabular}{|clrrr|}
	\toprule
	$R_{g+a}$ &         Player & g+a/90 &  $R_{vaep}$ & Market Value \\
	\midrule
	1 &      S. Agüero &            1.235 &          14 &        € 75m \\
	2 &       M. Salah &            1.232 &           2 &       € 150m \\
	3 &  P. Aubameyang &            1.191 &          42 &        € 75m \\
	4 &    P. Coutinho &            1.049 &           1 &       € 140m \\
	5 &    R. Sterling &            0.972 &           7 &       € 120m \\
	6 &        H. Kane &            0.905 &           9 &       € 150m \\
	7 &        L. Sané &            0.853 &          47 &       € 100m \\
	8 &       G. Jesus &            0.808 &         204 &        € 70m \\
	9 &     A. Martial &            0.795 &           6 &        € 60m \\
	10 &      O. Niasse &            0.749 &          17 &         € 7m \\
	\bottomrule
\end{tabular}
\end{subtable}
	\begin{subtable}{.48\textwidth}
	\centering
	\tabcolsep=0.13cm
	\vspace{.2cm}
	\caption{Top-10 players in terms of our \frameworkname player ratings}
	\label{tbl:pr-d}
\begin{tabular}{|clrrrrr|}
	\toprule
	$R_{vaep}$ &        Player & Rating &  $R_{g}$ &  $R_{a}$ &  $R_{g+a}$ & Market Value \\
	\midrule
	1 &   P. Coutinho &  0.899 &       10 &        2 &          4 &       € 140m \\
	2 &      M. Salah &  0.817 &        1 &       23 &          2 &       € 150m \\
	3 &  K. De Bruyne &  0.641 &       72 &        4 &         15 &       € 150m \\
	4 &     E. Hazard &  0.636 &       21 &      122 &         34 &       € 150m \\
	5 &     R. Mahrez &  0.635 &       34 &       11 &         16 &        € 60m \\
	6 &    A. Martial &  0.607 &       13 &       13 &          9 &        € 60m \\
	7 &   R. Sterling &  0.579 &        7 &        6 &          5 &       € 120m \\
	8 &      P. Pogba &  0.549 &       55 &        9 &         28 &        € 80m \\
	9 &       H. Kane &  0.545 &        4 &      140 &          6 &       € 150m \\
	10 &  S. Heung-Min &  0.539 &       19 &       36 &         17 &        € 50m \\
	\bottomrule
\end{tabular}
\end{subtable}
\end{table}

\subsection{Identifying promising young players and minor league talent}
\label{sec:talents}
\begin{table}
	\caption{The top-5 players born after January 1, 1997 in terms of our \frameworkname player ratings during the 2017/2018 season in (a) the tougher English and Spanish leagues, and (b) the smaller French, Dutch, and Belgian leagues.}
	\small
\begin{subtable}{.45\textwidth}
	\centering
	\tabcolsep=0.07cm
	\caption{Young talents in the English and Spanish leagues.}
	\label{tbl:talents2017}
	\begin{tabular}{|rllrrr|}
		\toprule
		Rank &                 Name &               Team &  Age & Rating & Market Value\\
		\midrule
		1 &          M. Rashford &  Man United &   20 &  0.406 & € 65m \\
		2 &  T. Alexander-Arnold &          Liverpool &   19 &  0.405 & € 45m\\
		3 &           O. Dembélé &          Barcelona &   20 &  0.360 & € 80m\\
		4 &             J. Kenny &            Everton &   21 &  0.344 & € 5m\\
		5 &      M. Oyarzabal &      Real Sociedad &   21 &  0.337 & € 40m\\
		\bottomrule
	\end{tabular}
\end{subtable}
\begin{subtable}{.45\textwidth}
	\centering
	\vspace{0.1cm}
	\caption{Young talents in the French, Dutch, and Belgian leagues.}
	\label{tbl:minor2017}
	\begin{tabular}{|rllrrr|}
		\toprule
		Rank &         Name &      Team &  Age & Rating & Market Value\\
		\midrule
		1 &  D. Neres &      Ajax &   21 &  0.620 & € 25m\\
		2 &     M. Mount &   Vitesse &   19 &  0.616 & € 4m\\
		3 &       Malcom &  Bordeaux &   21 &  0.567 & € 40m\\
		4 &    K. Mbappé &       PSG &   19 &  0.507 & € 200m\\
		5 &   F. de Jong &      Ajax &   20 &  0.495 & € 60m\\
		\bottomrule
	\end{tabular}
\end{subtable}

\end{table}

The English and Spanish leagues are the toughest and wealthiest by far. Hence, young players struggle to earn playing time, which forces the clubs to sign promising youngsters from smaller leagues such as the French, Dutch, and Belgian leagues. Typically, it is easier and especially cheaper for English and Spanish clubs to acquire promising youngsters from these leagues than from direct rivals. Therefore, we investigate the top-ranked young talents (i.e., players born after January 1, 1997 who played at least 900 minutes) separately for the 2017/2018 season in the English and Spanish leagues (Table~\ref{tbl:talents2017}), and the French, Dutch and Belgian leagues (Table~\ref{tbl:minor2017}).

% Hence, it is hard for young players to excel in them and English and Spanish clubs often look for promising young players in minor leagues such as the French, Dutch, and Belgian leagues, as those players are usually cheaper and easier to buy than promising young players from direct rivals. To account for this, we look at the top-5 young talents (players born after January 1st, 1997 who played at least 900 minutes) in the 2017/2018 season for the English and Spanish leagues (Table~\ref{tbl:talents2017}) and the French, Dutch, and Belgian leagues (Table~\ref{tbl:minor2017}) separately.

Marcus Rashford, who was linked with a € 110 million move to Real Madrid in January 2019,\footnote{https://www.thesun.co.uk/sport/football/8318008/real-madrid-marcus-rashford-transfer-man-utd/} and Ousmane Dembélé, who moved to Barcelona in August 2017 for a fee of € 120 million, are the most notable players in Table~\ref{tbl:talents2017}. In contrast, the fourth-ranked but lesser-known Jonjoe Kenny has a much lower estimated market value than both of these players due to two reasons. First, Kenny is a defensive player, who are typically valued lower than offensive players by clubs and fans. Second, Kenny plays for mid-table club Everton, where he is surrounded by only a few world-class players. Nevertheless, our player ratings suggest a much higher valuation than his current estimated market value of € 5 million.

% The most famous players in Table~\ref{tbl:talents2017} are Marcus Rashford, who was rumoured to transfer to Real Madrid for a fee of € 110 million in January 2019\footnote{https://www.thesun.co.uk/sport/football/8318008/real-madrid-marcus-rashford-transfer-man-utd/}, and Ousmane Dembélé, who transferred to Barcelona in summer 2017 for a fee of € 120 million. Jonjoe Kenny is a relatively unknown player and valued far less on the market than the other players. One explanation for this is that he is a defender, who are in general valued less by clubs and fans compared to attackers, at Everton, a mid-table club with few famous players. Given the value Jonjoe can bring to a team, we consider him a steal at a current market value of only € 5 million.

David Neres tops Table~\ref{tbl:minor2017}. In the summer of 2017, the winger became the fourth most expensive incoming transfer in the Dutch league when Ajax acquired him for € 15 million.
%Top clubs Liverpool, Chelsea, and Arsenal have all expressed interest in signing him in summer 2019.
He is now a transfer target for top clubs Liverpool, Chelsea, and Arsenal, who all wish to sign him in the summer of 2019.
Second-ranked Mason Mount was with Dutch side Vitesse on a season-long loan from Chelsea and received Vitesse's Player of the Year Award. 
%He received Vitesse's Player of the Year award for his contributions to the club's play-off qualification.
%Although our player ratings suggest Mount is considerably undervalued with an estimated valuation of € 4 million, the midfielder failed to break into Chelsea's first team and was sent on another season-long loan.
Fourth-ranked Kylian Mbappé won the Best Young Player Award at the 2018 World Cup, while both Malcom (summer 2018, fee € 41 million) and Frenkie de Jong (summer 2019, fee € 75 million) have signed with Barcelona.

% David Neres tops Table~\ref{tbl:minor2017} as an essential forward at Ajax, who bought him in summer 2017 from São Paulo FC for a transfer fee of € 15 million, making him the fourth highest incoming transfer ever in the Dutch League. Mason Mount is a Chelsea player who joined Dutch club Vitesse on a loan in season 2017/2018. He won the Vitesse Player Of The Year award and was instrumental in guiding Vitesse to the Dutch league play-offs. His current market value is rather low at € 4 million and he has been put on loan again in the 2018/2019 season rather than make it to Chelsea's first team. However, given Chelsea's reputation for not giving opportunities to promising young player (e.g., Lukaku, De Bruyne, Salah), we suspect that he is currently undervalued. Kylian Mbapp\'{e} is recognized as one of the biggest talents in the world and was a break-out star at the 2018 World Cup. Both Malcom (summer 2018, fee € 41 million) and Frenkie de Jong (summer 2019, fee € 75 million) have transferred to Barcelona.

Tables~\ref{tbl:talents2017}~and~\ref{tbl:minor2017} demonstrate our framework's ability to serve as a useful tool for talent scouts. Our framework can generate rankings for each league in the world (e.g., second divisions or leagues in North America, South America, and Asia) given that the required event stream data is available.

% Tables \ref{tbl:talents2017} and \ref{tbl:minor2017} were generated fully automatically. This means that, given the correct event stream data,  we can use our framework to instantly generate this type of ranking for other minor leagues (e.g., second divisions, leagues in North America, South America, and Asia), which would be a helpful tool for scouts.

\subsection{Characterizing playing style}

Clubs are increasingly considering player styles during the recruitment process to identify players who best suit their team's preferred style of play (e.g., short passes and high defending vs. long balls and defensive play). Currently, scouts are typically tasked with judging playing styles with the naked eye. However, these scouts' time is often the limiting resource, which makes it difficult to consider the entire pool of candidate reinforcements. Therefore, metrics that assess a player's ability to perform different types of actions can help select a relevant set of players who are worth extra attention. Using our \frameworkname framework, addressing this task boils down to computing a player's rating per 90 minutes for each type of action.

\begin{figure}
	\begin{subfigure}{0.45\textwidth}
	\centering
	\includegraphics[width=\textwidth]{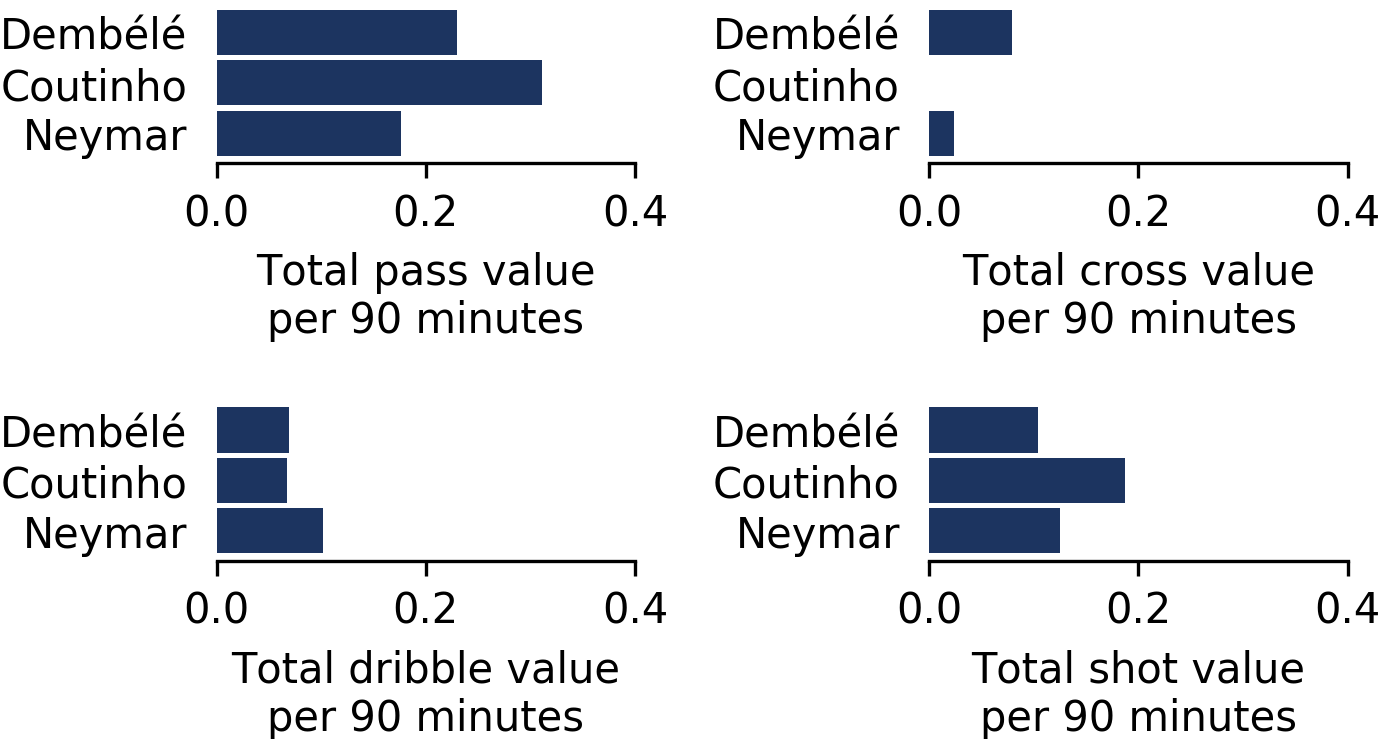}
	\caption{Playing style of replacement players for Neymar}
	\label{fig:neymar}
	\end{subfigure}
	\begin{subfigure}{0.45\textwidth}
	\centering
	\vspace{0.2cm}
	\includegraphics[width=\textwidth]{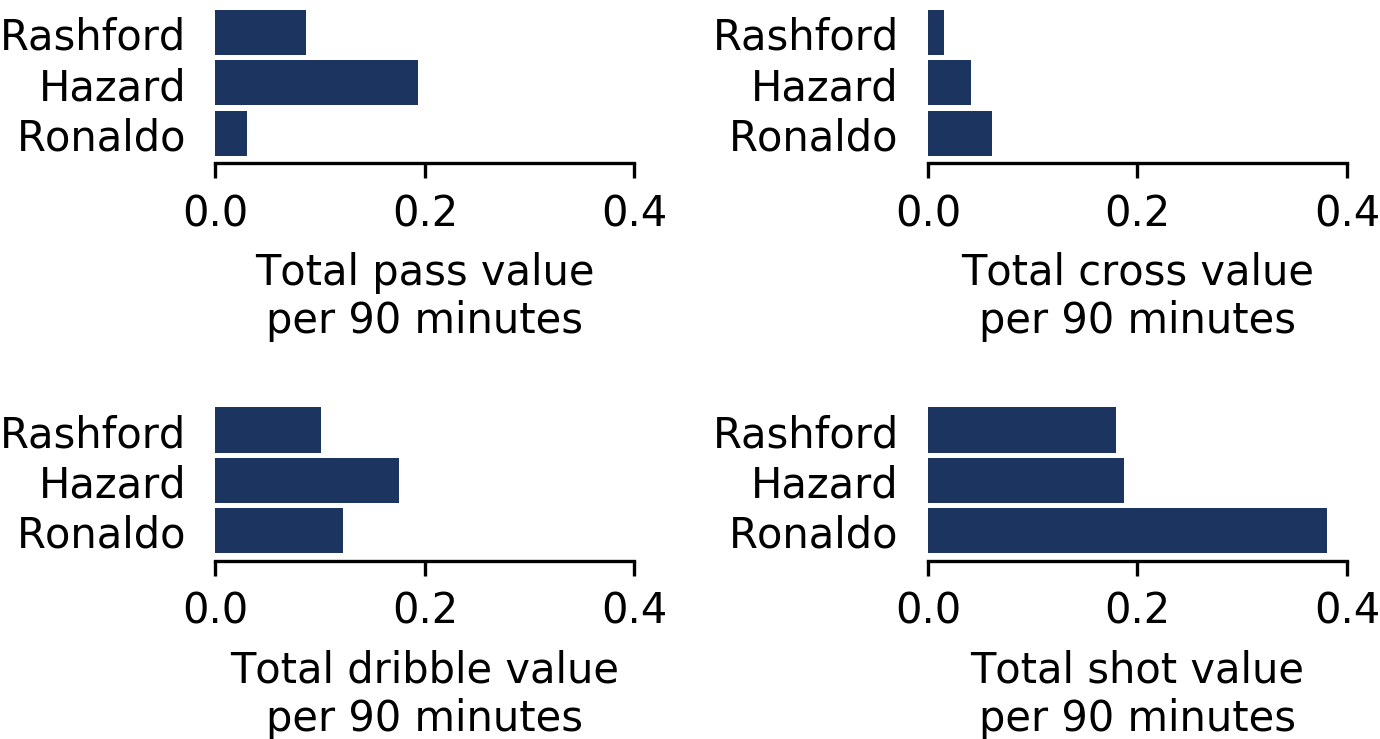}
	\caption{Playing style of replacement players for Ronaldo}
	\label{fig:ronaldo}
	\end{subfigure}
	\caption{Overview of the total contribution per 90 minutes for different types of actions for (a) Neymar, Ousmane Demb\'el\'e, and Philippe Coutinho during the 2016/2017 season, and (b) Cristiano Ronaldo, Marcus Rashford, and Eden Hazard during the 2017/2018 season.}
\end{figure}

As a concrete use case, consider Barcelona's attempts in the summer of 2017 to offset the loss of Neymar by acquiring Borussia Dortmund's Ousmane Demb\'{e}l\'{e} and Liverpool's Philippe Coutinho. Figure~\ref{fig:neymar} compares Demb\'{e}l\'{e},  Coutinho and Neymar's total ratings per 90 minutes for four action types. According to our metric, both Demb\'{e}l\'{e} and Coutinho's passes receive a higher value than Neymar's while Neymar is a superior dribbler. %, but Demb\'{e}l\'{e} and Coutinho also perform decent.
From a stylistic perspective, this breakdown suggests that both Demb\'{e}l\'{e} and Coutinho were reasonable targets as not many players come close to replicating Neymar's signature skill of dribbling. Demb\'{e}l\'{e} and Coutinho are decent dribblers and better passers than Neymar. In addition, Demb\'{e}l\'{e} outperforms Neymar in crossing, while Coutinho outperforms him in shooting.

Similarly, Real Madrid lost their all-time top scorer Cristiano Ronaldo in the summer of 2018. The struggling club appears in desperate need of a suitable replacement. Manchester United's Marcus Rashford and Chelsea's Eden Hazard have both been linked with moves to Madrid. However, Figure~\ref{fig:ronaldo} shows that neither comes close to replicating Ronaldo's incredible finishing skill. Moreover, Ronaldo exhibits a higher total shot value per 90 minutes than Rashford and Hazard combined. While Hazard outperforms Rashford in every aspect, Rashford is closer to Ronaldo in terms of style as both rate similarly for passing and dribbling. If Real Madrid want to stick to their current playing style, our analysis suggests that the 21-year-old Rashford would be the better choice. However, if their aim is to immediately strengthen their team, then the 28-year-old Hazard would be the preferred choice as he is a better player regardless of his specific playing style.

% If Real Madrid want to stick to their current playing style, our analysis suggests that the 21-year-old Rashford would be the better choice. However, if their aim is to immediately strengthen their team, then the 28-year-old Hazard would be the preferred choice as he is a better player regardless of his specific playing style.

% Two players they have been trying to acquire are Manchester United's Marcus Rashford and Chelsea's Eden Hazard. As can be seen in Figure \ref{fig:ronaldo}, neither comes close to replicating Ronaldo's incredible finishing skill; Ronaldo contributes almost double the total shot value per 90 minutes of Rashford and Hazard. While Hazard seems to be a better player than Rashford in every aspect, Rashford is closer in style to Ronaldo, as they both have similar (lower) values for passes and dribbles. If Real Madrid's goal is to build the new Ronaldo in the long term, we recommend them to acquire Rashford, as he is only 21 years old and stylistically comes closer to Ronaldo. If their goal is to immediately strengthen their team with a skilled attacker (regardless of style), we recommend Hazard, as he is currently a better player than Rashford, but has a different playing style from Ronaldo and is already 28 years old.

\subsection{Trading off action quality and quantity}
\begin{figure*}
	\begin{subfigure}{0.31\textwidth}
		%\centering
		%\includegraphics[width=\textwidth]{wyscout/multi-tradeoff-2017.png}
		%\includegraphics[width=\textwidth]{wyscout/trade-off-all-fixed.pdf}
		\includegraphics[width=\textwidth]{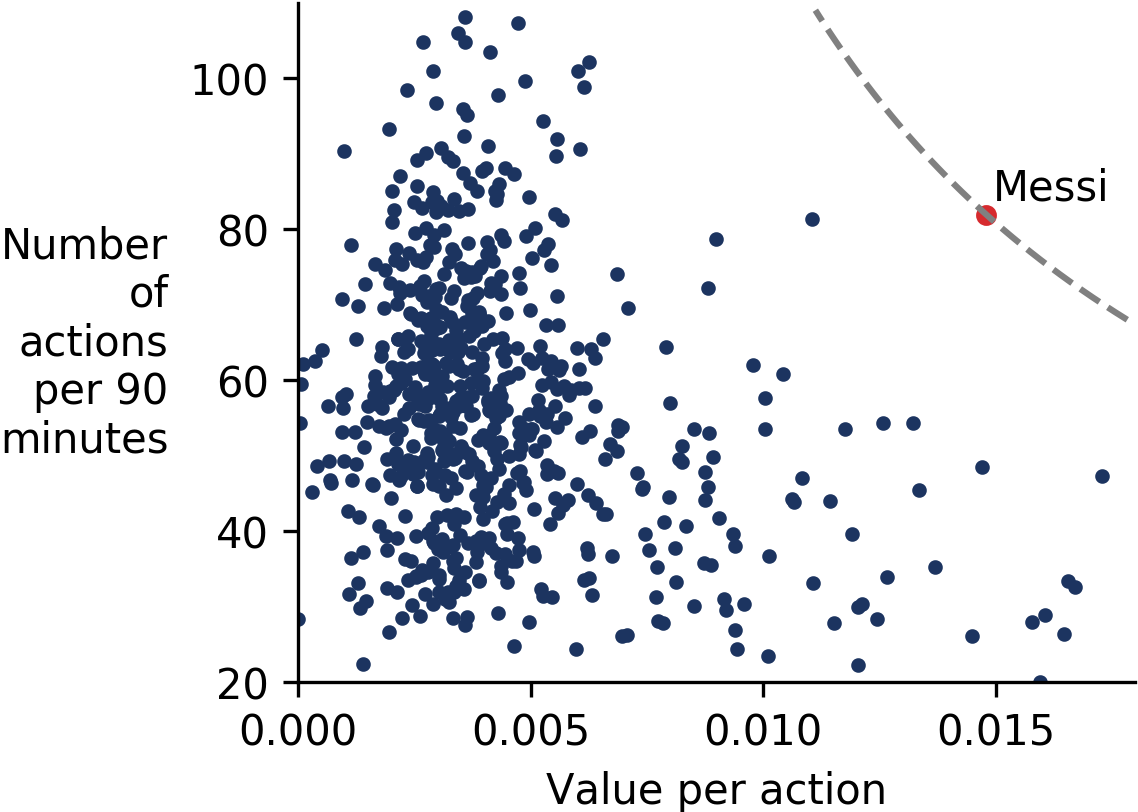}
		\caption{All players}
		\label{fig:trade-off}
	\end{subfigure}
	\hfill
	\begin{subfigure}{0.31\textwidth}
	\includegraphics[width=\textwidth]{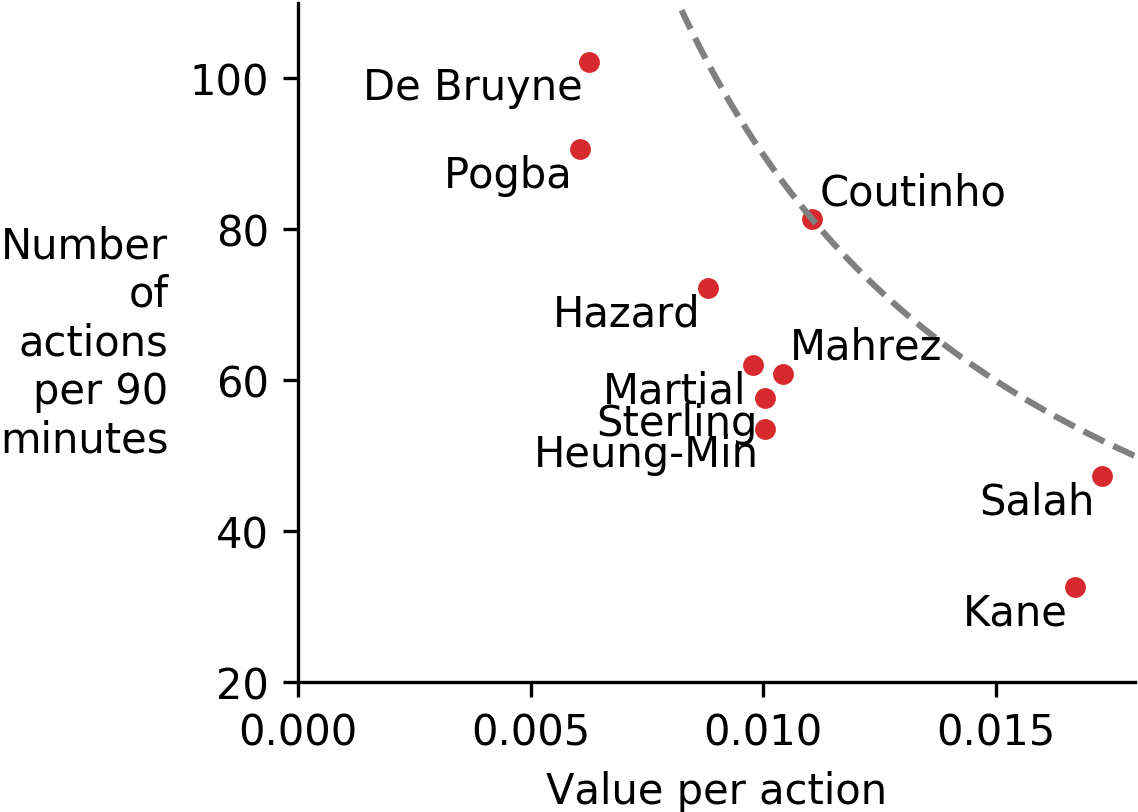}
	\caption{Top-10 players in the English league}
	\label{fig:english-tradeoff}
	\end{subfigure}
	\hfill
	\begin{subfigure}{0.33\textwidth}
	\includegraphics[width=\textwidth]{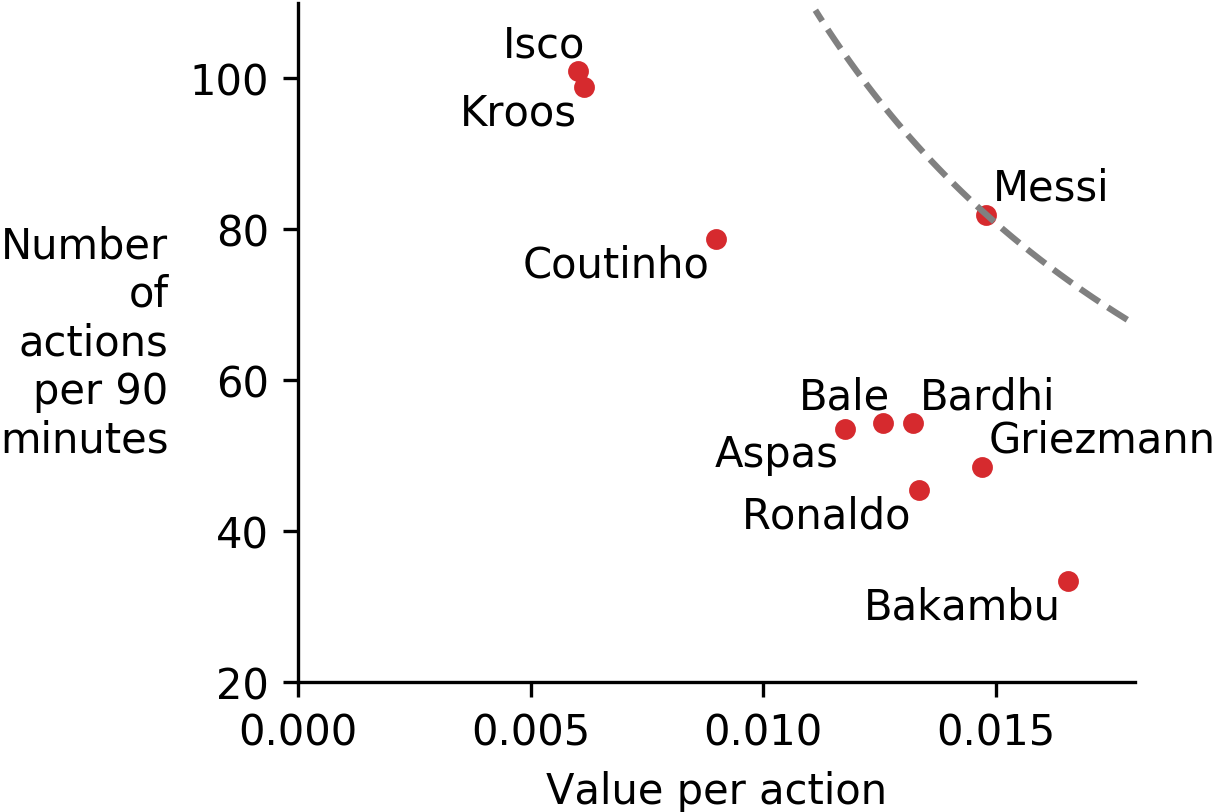}
	\caption{Top-10 players in the Spanish league}
	\label{fig:spanish-tradeoff}
	\end{subfigure}
	\caption{Scatter plots of players in the 2017/2018 season who played at least 900 minutes in the Spanish or English league. The plots contrast the average number of actions performed per 90 minutes with the average value of the actions of the player. As shown by the grey-dotted isoline in (a) and (c), Lionel Messi is clearly in a class of his own.}
\end{figure*}

A natural tension exists between the quality and quantity of actions.
If a player performs a high number of actions, then it is harder for each action to have a high value.
Figure~\ref{fig:trade-off} shows the number of actions that players execute on average per 90 minutes (quantity) and the average value of these actions (quality) for those players who played at least 900 minutes during the 2017/2018 season in the Spanish and English leagues. 
The grey-dotted isoline shows the gap in \frameworkname rating between top-ranked Lionel Messi and the rest. The isoline is curved as a player's rating is obtained by multiplying the average value per action (\emph{x-axis}) and the average number of actions (\emph{y-axis}). As shown by the isoline and more traditional statistics,\footnote{\url{https://fivethirtyeight.com/features/lionel-messi-is-impossible/}} Messi is clearly in a class of his own.

Zooming in on Figure~\ref{fig:trade-off}, Figure~\ref{fig:english-tradeoff} shows the top-10 players in the 2017/2018 English Premier League season. Strikers Harry Kane and Mohammed Salah perform a relatively low number of actions but their actions are highly valued on average. Midfielders Kevin De Bruyne and Paul Pogba perform more actions albeit with a lower average value per action. Philippe Coutinho, Eden Hazard, Riyad Mahrez, Anthony Martial, Raheem Sterling, and Son Heung-min fall in between these two archetypes, hitting a sweet spot between the quality and quantity of their actions.

% Figure~\ref{fig:english-tradeoff} zooms in on Figure \ref{fig:trade-off} and shows only the top-10 players in the English Premier League during the 2017/2018 season. Strikers Harry Kane and Mohammed Salah perform a relatively low number of actions, but each action is highly valued. Midfielders Kevin De Bruyne and Paul Pogba perform more actions albeit with a lower average value per actions. Philippe Coutinho, Eden Hazard, Riyad Mahrez, Anthony Martial, Raheem Sterling, and Son Heung-min fall in between these two archetypes, hitting a sweet spot between action quality and quantity.

%\begin{figure}
%	\centering
%	\includegraphics[width=.40\textwidth]{wyscout/english-tradeoff-2017.png}
%	\caption{Scatter plot that contrasts the average number of actions performed per 90 minutes with the average value of these actions for the top-10 players in the English Premier League during the 2017/2018 season}
%	\label{fig:english-tradeoff}
%\end{figure}

Similarly, Figure~\ref{fig:spanish-tradeoff} shows the top-10 players in the Spanish league. We observe the same archetypes as for the English league. Strikers Cristiano Ronaldo, Antoine Griezmann, Gareth Bale, Enis Bardhi, Iago Aspas, and C\'edric Bakambu perform a low number of highly valuable actions. Real Madrid midfielders Toni Kroos and Isco perform more actions that are less valuable. Philippe Coutinho, who appears in both figures following his move from Liverpool to Barcelona in January 2018, again hits the sweet spot between action quality and quantity. Lionel Messi is an outlier in the sense that he rates high on action quality and quantity at the same time.

%\begin{figure}
%	\centering
%	\includegraphics[width=.40\textwidth]{wyscout/spanish-tradeoff-2017.png}
%	\caption{Scatter plot that contrasts the average number of actions performed per 90 minutes with the average value of these actions for the top-10 players in the Spanish Premier League during the 2017/2018 season}
%	\label{fig:spanish-tradeoff}
%\end{figure}

\subsection{Evaluating design choices}
\label{sec:design-choices}

A common challenge in data science is to evaluate the performance of a system. While it can be easy to define a high-level task such as assigning values to actions, one underappreciated aspect of evaluating the solution is that usually no ground truth exists and standard evaluation metrics such as accuracy, precision and recall thus cannot be used. As a result, the only way to evaluate a system is to evaluate the components it consists of. In our case, we evaluate the action values by evaluating the underlying scoring and conceding probabilities for which ground truth labels are available.

% One underappreciated aspect of this is that while it can be easy to define a high-level task such as assigning values to actions, oftentimes no ground truth exists for the task. This means that it is impossible to compute standard evaluation metrics such as accuracy, precision or recall. This necessitates identifying other variables that serve as proxies for the results of our system. In our case, the scoring and conceding probabilities make a good proxy for our action values because they directly influence the action values and there are ground truth labels available for the scoring and conceding probabilities.

\subsubsection{Evaluation methodology}

To generate scoring and conceding probabilities, we train classification models using the features described in Section~\ref{sec:algorithm} with the CatBoost algorithm. We evaluate these design choices by comparing the performance of our approach to alternative approaches that use either a different feature set or a different algorithm.

Similar to our main approach, we train two classification models for each alternative: we train a first model on the 2012/2013 through 2015/2016 seasons to produce the outcomes for the 2016/2017 season, and a second model on the 2012/2013 through 2016/2017 seasons to produce the outcomes for the 2017/2018 season.

We evaluate the performance of each approach using two metrics often used for evaluating probabilistic predictions: the Brier score and ROC AUC \cite{ferri2009experimental}. The Brier score measures the accuracy and calibration of the predictions and is minimized when the true underlying probability distribution of the data is reported. This property is important because we sum and subtract the predicted probabilities to generate action values. 
The area under the receiver operator curve (ROC AUC) evaluates how well the approaches can discern positive examples from negative examples. An important advantage of ROC AUC is that the metric is unaffected by unbalanced data sets, as in our data only $1.5\%$ ($0.5\%$) of all game states lead to a scored (conceded) goal.

\subsubsection{Choice of feature set}

Most existing models that analyze soccer event data only use location and action type data~\cite{altman2015beyond,decroos2017predicting}. To evaluate the usefulness of our comprehensive feature set (detailed in Section~\ref{sec:algorithm}), we compare it to four baseline feature sets: no features,\footnote{
	The \emph{no features} baseline always predicts the mean class probability, i.e., if our data set contains 1.5\% positive examples, we always predict 0.015.
} location, action type, and location + action type (Table~\ref{tbl:design-scores}). 
We evaluate each feature set using the CatBoost algorithm.
For estimating both $P_{scores}$ and $P_{concedes}$, our feature set outperforms the baseline feature sets on both evaluation metrics. This result suggests that our features capture important game state context that is missing from the baseline feature sets.

\subsubsection{Choice of learning algorithm}
\label{sec:choice-of-learning-algorithm}

The most popular choices of learning algorithm in data science projects are Logistic Regression~\cite{pedregosa2011scikit}, Random Forest~\cite{pedregosa2011scikit}, and more recently XGBoost~\cite{chen2016xgboost} and CatBoost~\cite{prokhorenkova2018catboost}. Gradient boosting methods have a successful track record in a variety of learning problems with heterogeneous features, noisy data, and complex dependencies.
Table~\ref{tbl:design-scores} compares the performances of the four learning algorithms using the features from Section~\ref{sec:algorithm}. CatBoost performs best in all cases, with XGBoost a close second. This narrow victory can be attributed to CatBoost's intelligent handling of categorical features compared to XGBoost's more naive one-hot encoding.

%\subsubsection{Choice of number of actions to consider}
%
%We investigate the optimal number of previous actions to construct the game states. Considering too few actions might leave valuable contextual information unused, while considering too many actions can make the feature set unnecessarily noisy. We trained five different models using the CatBoost algorithm varying the number of previous actions from one through five (Table~\ref{tbl:design-scores}). Our experiment shows that three is the optimal number of actions. Including more than three actions does not improve the performance and even degrades the performance again in the case of $P_{concedes}$. 

\begin{table}
	\caption{Different design choices evaluated on both scoring and conceding probabilities using the Brier score and ROC AUC.  For the Brier score lower values are better, whereas for ROC AUC higher values are better.}
	\small
	\tabcolsep=0.12cm
	\begin{tabular}{|lrrrrr|}
		\toprule
		& & \multicolumn{2}{c}{$P_{scores}$} & \multicolumn{2}{c|}{$P_{concedes}$}\\
		%\cmidrule(r){3-4} \cmidrule(l){5-6}
		\textbf{Design} & \textbf{Choice} & \textbf{Brier} & \textbf{AUC} &  \textbf{Brier} & \textbf{AUC}\\
		\midrule
		Feature set & All features & \textbf{0.01376 } & \textbf{0.7693} & \textbf{0.00547} & \textbf{0.7313}\\
		& No features &  0.01632  & -  & 0.00564 & -\\
		& Location & 0.01562 &0.7330 & 0.00560  & 0.6770\\
		& Action type & 0.01590 & 0.6405 & 0.00562 & 0.6348\\
		& Loc + Action type& 0.01549 & 0.7417 & 0.00550  &0.6912\\
		\midrule
		Algorithm & CatBoost &  \textbf{0.01376 } & \textbf{0.7693}  & \textbf{0.00547} & \textbf{0.7313}\\
		& Logistic Regression & 0.01601  & 0.7231& 0.00562  & 0.6578\\
		& Random Forest & 0.01409  & 0.7050& 0.00552  & 0.6457\\
		& XGBoost & 0.01390  & 0.7556& 0.00550 & 0.7255\\
%		\midrule
%		Actions & 1 action & 0.01378  & 0.7657 & 0.00557  & 0.6866\\
%		& 2 actions & 0.01376 & 0.7687 & 0.00548  & 0.7301 \\
%		& 3 actions &  \textbf{0.01376 } & \textbf{0.7693} &\textbf{0.00547} & \textbf{0.7313}\\
%		& 4 actions& 0.01376 &0.7690 & 0.00556  & 0.7130\\
%		& 5 actions & 0.01376 &0.7692 & 0.00565  & 0.6951\\
		\bottomrule
	\end{tabular}
	\label{tbl:design-scores}
\end{table}

\subsection{Discussion of remaining challenges}

One limitation of our \frameworkname framework is that we only value on-the-ball actions. That is, the model only values actions with the ball, while defending is often more about preventing your opponent from gaining possession of the ball by clever positioning and anticipation.

Another challenge is that it is hard to accurately compare players across leagues, as it is easier to perform highly valuable actions in minor leagues (e.g., French, Dutch, and Belgian) than in tougher leagues (e.g., English and Spanish). This can be clearly observed in Section \ref{sec:talents} where the young talents in the minor leagues receive a higher rating than those in the English and Spanish leagues.

Similarly, it can even be hard to accurately compare players across clubs in the same league as it is generally easier to perform valuable actions in a top club with strong teammates, than in a mid-table club with weaker teammates.

The final challenge for deploying our framework in the real world is building trust in the ratings as traditional scouts are unfamiliar with our way of rating soccer players. In addition, our ratings are slightly less intuitive than traditional metrics such as goals per 90 minutes, which complicates the task for analytically less inclined scouts to understand what our ratings measure precisely.

%The SciSports Datascouting department leverages our action values for providing data-driven advice to soccer clubs and associations for player recruitment and opponent analysis. Until recently, these datascouts almost exclusively relied upon more traditional metrics and statistics. While our action values are currently only available for internal use by the datascouts, they will shortly also appear in the SciSports Insight %\footnote{\url{https://insight.scisports.com}} 
%online scouting platform. 

\section{Related work}
\label{sec:related-work}

While valuing player actions in soccer is an important task, it has remained virtually unexplored due to the challenges resulting from the dynamic and low-scoring nature of soccer. The approaches from \citet{norstebo2016valuing}, \citet{bransen2019pressure} and \citet{fernandez2019decomposing} for soccer, \citet{routley:uai15} and~\citet{liu:ijcai18} for ice hockey, and \citet{cervone2014pointwise} for basketball come closest to our framework. Most of these approaches address the task of valuing individual actions by modeling a game as a Markov game~\citep{littman1994markov}. In contrast to \citet{norstebo2016valuing} and \citet{routley:uai15}, which divide the pitch into a fixed number of zones, our approach models the exact locations of each action. Unlike \citet{cervone2014pointwise}, which values only three types of on-the-ball actions, our approach considers any relevant on-the-ball action during a game. However, our definitions of player actions, game states, and action values are similar to those used by these works as well as earlier research for soccer~\citep{rudd2011framework, hirotsu2002using}, American football~\citep{goldner2012markov}, and baseball~\citep{tango2007book}.

Most of the related work on soccer either focuses on a limited number of player action types like passes and shots or fails to account for the circumstances under which the actions occurred. \citet{decroos2017starss}, \citet{knutson2017introducing}, and \citet{gregory2017how} address the task of valuing the actions leading up to a goal attempt, whereas \citet{bransen2019measuring}, \citet{bransen2018mlsa}, and \citet{gyarmati2016qpass} address the task of valuing individual passes. The former approaches naively assign credit to the individual actions by accounting for a limited amount of contextual information only, while the latter approaches are limited to a single type of action.

Furthermore, this work is also related to expected-goals models, which estimate the probability of a goal attempt resulting into a goal \citep{lucey2014quality,caley2015premier,altman2015beyond,mackay2017predicting,decroos2017predicting}. In our \frameworkname framework, computing the expected-goals value of a goal attempt boils down to estimating the value of the game state prior to the goal attempt.

\section{Conclusion}
\label{sec:conclusion}
This paper introduced \repname, a language for representing event stream data that is designed with the goal of facilitating data analysis, and \frameworkname, a framework for assigning a value to each individual player action during a soccer game. The advantages of \frameworkname over most existing works are that it (1) values all action types (e.g., passes, crosses, dribbles, and shots), (2) bases its valuation on the game context, and (3) reasons about an action's possible effects on the subsequent actions. Intuitively, the player actions that increase a team's chance of scoring receive positive values while those actions that decrease a team's chance of scoring receive negative values.

\section{Acknowledgments}
Tom Decroos is supported by the Research Foundation-Flanders (FWO-Vlaanderen). Jesse Davis is partially supported by the EU Interreg VA project Nano4Sports and the KU Leuven Research Fund (C14/17/07, C32/17/036). The authors thank Wyscout for supplying the event stream data used in this paper.

\bibliographystyle{ACM-Reference-Format}
%\bibliography{references.bib} %linux
\bibliography{references} %windows
\clearpage
\appendix

\section{Appendix on reproducibility}
\subsection{\repname action types}
\label{app:action-types}

Table~\ref{tbl:action-types} provides an overview of the 21 action types in the \repname representation alongside their descriptions.

\subsection{Data description}

\begin{figure}[H]
	\centering
	\includegraphics[width=0.45\textwidth]{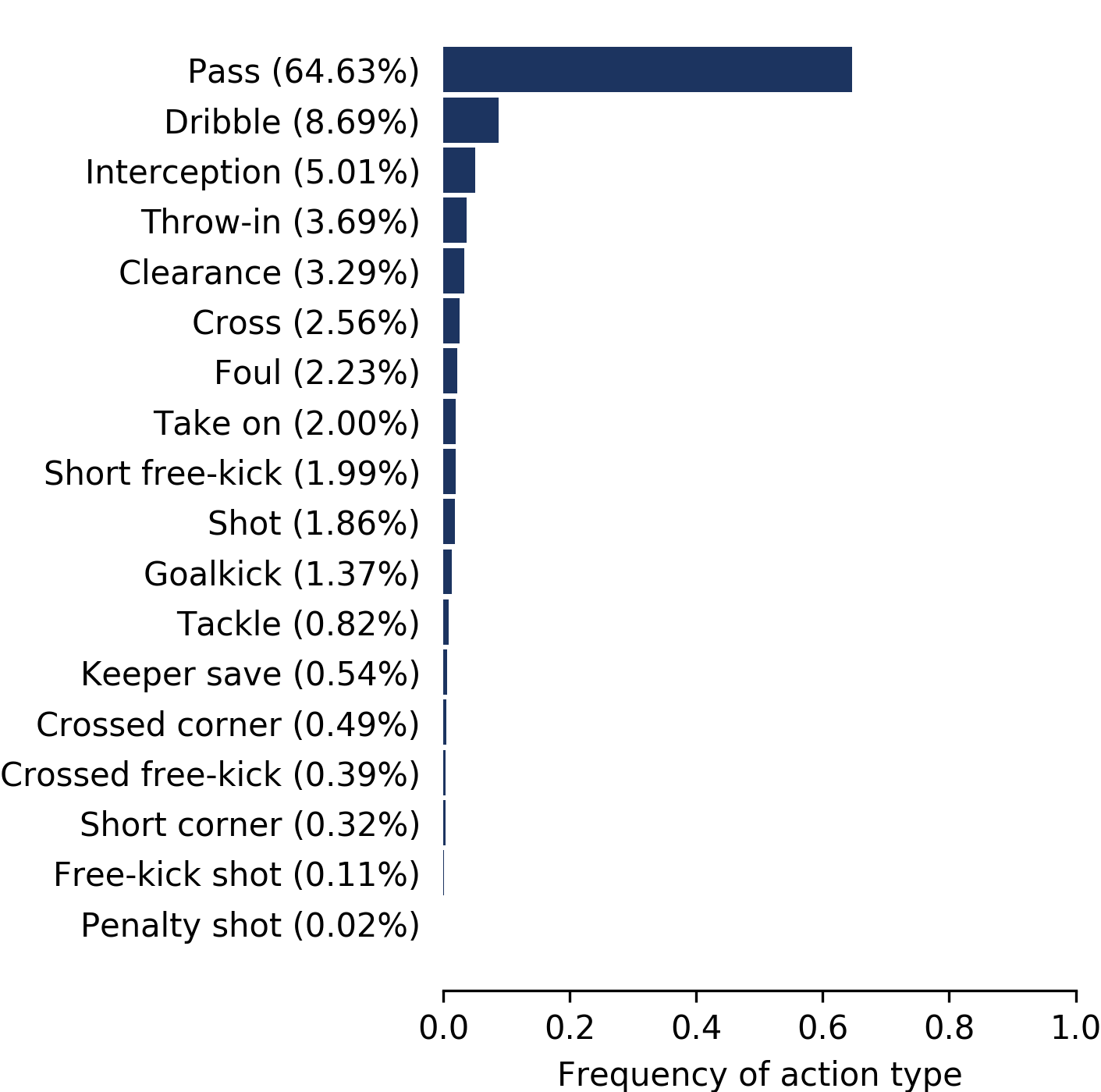}
	\caption{Frequency of each action type in the Wyscout data converted to the \repname representation.}
	\label{fig:frequency}
\end{figure}

We ran our experiments on Wyscout data for the English, Spanish, German, Italian, French, Dutch, and Belgian top divisions. We considered 11,565 games played in the 2012/2013 through 2017/2018 seasons. After transforming the Wyscout data to our \repname representation, each game contains $\pm$1250 actions on average. As shown in Figure~\ref{fig:frequency}, the most frequent action types in our data set are passes (64.63\%), dribbles (8.69\%), and interceptions (5.01\%).

\subsection{Experimental setup and implementation}

We performed all experiments in this paper in Python. We evaluated the performance of four popular learning algorithms:

 \begin{description} 
	\item[Logistic Regression] We used the implementation from the \texttt{scikit-learn}\footnote{https://scikit-learn.org/stable/modules/generated/sklearn.linear\_model.LogisticRegression.html} Python package. We used an L2 regularization penalty and L-BFGS as the solver for the optimization problem.
\item[Random forest] We used the implementation from the \texttt{scikit\-learn}\footnote{https://scikit-learn.org/stable/modules/generated/sklearn.ensemble.RandomForestClassifier.html} Python package. We trained a forest of 100 trees using 40 parallel threads.
\item[XGBoost] We used the official Python implementation.\footnote{https://xgboost.ai} We trained 100 trees of maximum depth 3 with a learning rate of 0.1 using 40 parallel threads.
\item[CatBoost] We used the official Python implementation from Yandex.\footnote{https://tech.yandex.com/catboost/} We set all parameters to their default values, except for the number of parallel threads, which we set to 40.
 \end{description} 

We trained and evaluated all models on a computing server running Ubuntu 16.04 with 128GB of RAM and two CPUs of type Xeon(R) CPU E5-2630 v4 @ 2.20GHz, providing a maximum of 20 cores and 40 threads. The runtime for each learning algorithm per task is available in Table~\ref{tbl:runtimes}.

\begin{table*}
	\caption{Overview of the 21 action types in \repname alongside their descriptions. The \textit{Success?} column specifies the condition the action needs to fulfil to be considered successful, while the \textit{Special} column lists additional possible result values.}
	\label{tbl:action-types}
	\begin{tabular}{|llll|}
		\toprule
		\textbf{Action type} & \textbf{Description} & \textbf{Success?} & \textbf{Special result} \tabularnewline\midrule
		Pass & Normal pass in open play & Reaches teammate & Offside \tabularnewline%\midrule
		Cross & Cross into the box & Reaches teammate & Offside \tabularnewline%\midrule
		Throw-in & Throw-in & Reaches teammate  & - \tabularnewline%\midrule
		Crossed corner & Corner crossed into the box & Reaches teammate & Offside \tabularnewline%\midrule
		Short corner & Short corner  & Reaches teammate & Offside \tabularnewline%\midrule
		Crossed free-kick & Free kick crossed into the box & Reaches teammate & Offside \tabularnewline%\midrule
		Short free-kick & Short free-kick & Reaches team mate & Offside \tabularnewline%\midrule
		Take on & Attempt to dribble past opponent & Keeps possession & - \tabularnewline%\midrule
		Foul & Foul & Always fail & Red or yellow card \tabularnewline%\midrule
		Tackle & Tackle on the ball & Regains possession & Red or yellow card \tabularnewline%\midrule
		Interception & Interception of the ball & Always success & - \tabularnewline%\midrule
		Shot & Shot attempt not from penalty or free-kick & Goal & Own goal \tabularnewline%\midrule
		Penalty shot & Penalty shot & Goal & Own goal \tabularnewline%\midrule
		Free-kick shot & Direct free-kick on goal & Goal & Own goal \tabularnewline%\midrule
		Keeper save & Keeper saves a shot on goal  & Always success & - \tabularnewline%\midrule
		Keeper claim& Keeper catches a cross   & Does not drop the ball & - \tabularnewline%\midrule
		Keeper punch& Keeper punches the ball clear  & Always success & - \tabularnewline%\midrule
		Keeper pick-up& Keeper picks up the ball & Always success & - \tabularnewline%\midrule
		Clearance & Player clearance  & Always success & - \tabularnewline%\midrule
		Bad touch & Player makes a bad touch and loses the ball & Always fail & - \tabularnewline%\midrule
		Dribble & Player dribbles at least 3 meters with the ball & Always success & - \tabularnewline
		%Run without ball & Player runs without the ball & Always success & - \tabularnewline
		\bottomrule
	\end{tabular}
\end{table*}

\begin{table*}
	%\small
	\centering
	\caption{Runtimes for each learning algorithm per task. The two training sets, seasons 2012/2013 through 2015/2016 and seasons 2012/13 through 2016/2017, contain respectively 8,518,378 and 11,438,956 actions each. The two evaluation sets, season 2016/2017 and season 2017/208, contain respectively 2,920,578 and 2,988,847 actions each.}
	\label{tbl:runtimes}
	\begin{tabular}{|lrrrr|}
		\toprule
		Task & Logistic Regression & Random Forest & XGBoost & CatBooost\\
		\midrule
		Training on seasons 2012/2013 - 2015/2016 &4 minutes& 16 minutes& 16 minutes& 100 minutes\\
		Training on seasons 2012/2013 - 2016/2017 & 6 minutes&25 minutes&22 minutes&140 minutes\\
		Predicting on season 2016/2017 &30 seconds&1 minute & 40 seconds&3 minutes\\
		Predicting on season 2017/2018 &30 seconds&1 minute&50 seconds&3 minutes\\
		\bottomrule
	\end{tabular}
\end{table*}

\end{document}